\documentclass[12pt]{article}

\usepackage{epsfig,latexsym,amsfonts,amsmath,amsthm,amssymb,amsbsy,multirow,slashed,wasysym,textcomp,subfigure,hyperref,wrapfig,datetime,comment,mathtools,cancel,cite,mathrsfs,footmisc,mathrsfs}

\usepackage{scalerel}
\hypersetup{colorlinks=true, linkcolor=green, citecolor=cyan, linktoc=page}
\makeatother

\usepackage{stackengine,wasysym}
\usepackage{booktabs}

\usepackage[font=small,skip=-15pt]{caption} 
\def\bee#1\eee{\begin{align}#1\end{align}}

\def\ee{\end{equation}}
\def\bea{\begin{eqnarray}}
\def\eea{\end{eqnarray}}
\newcommand{\beq}{\begin{eqnarray}}
\newcommand{\eqq}{\end{eqnarray}}
 \newcommand{\badat}{\begin{alignedat}}
 \newcommand{\eadat}{\end{alignedat}}

\newcommand{\eal}[1]{\be \begin{aligned} #1 \end{aligned}\end{equation}}
\newcommand{\eqn}[1]{\be #1 \end{equation}}
\newcommand{\eqa}[1]{\bea  #1\end{eqnarray}}

\long\def\new#1\endnew{{\bf #1}}		
\long\def\del#1\enddel{}

\def\del{\partial}

\usepackage{array}
\usepackage{tikz}
\usepackage{multirow,multicol}
\usepackage{xcolor}
\usepackage{dirtytalk}
\usepackage{mathrsfs}
\definecolor{oldmauve}{rgb}{0.4, 0.19, 0.28}
\definecolor{pansypurple}{rgb}{0.47, 0.09, 0.29}
\definecolor{burgundy}{rgb}{0.5, 0.0, 0.13}
\definecolor{carminepink}{rgb}{0.92, 0.3, 0.26}
\definecolor{blue(pigment)}{rgb}{0.2, 0.2, 0.6}
\definecolor{darkseagreen}{rgb}{0.56, 0.74, 0.56}
\definecolor{darkspringgreen}{rgb}{0.09, 0.45, 0.27}
\definecolor{ceruleanblue}{rgb}{0.16, 0.32, 0.75}

 \usepackage[framemethod=tikz]{mdframed}
\definecolor{mycolor}{rgb}{0.122, 0.435, 0.698}

\newmdenv[innerlinewidth=0.5pt, roundcorner=4pt,linecolor=mycolor,innerleftmargin=6pt,
innerrightmargin=6pt,innertopmargin=6pt,innerbottommargin=6pt]{bluebox}

\newcommand{\be}{\begin{eqnarray}}
\newcommand{\DD}{\mathring{\Delta}}
\newcommand{\en}{\end{eqnarray}}

\numberwithin{equation}{section} 

\begin{document}

\begin{titlepage}
  \thispagestyle{empty}

  \begin{center}


{\Large\textbf{Universal Relations with the Non-Extensive Entropy Perspective}}\\
\vspace{0.2cm}

\vskip1cm
Ankit Anand$^\star$\footnote{\fontsize{8pt}{10pt}\selectfont\ \href{mailto:ankitanandp94@gmail.com}{ankitanandp94@gmail.com}} and Saeed Noori Gashti $^{\dagger}$\footnote{\fontsize{8pt}{10pt}\selectfont\ \href{mailto:saeed.noorigashti@stu.umz.ac.ir}{saeed.noorigashti@stu.umz.ac.ir}}

\vskip0.5cm

\normalsize
\medskip

$^\star$\textit{Physics Division, School of Basic and Applied Sciences, Galgotias University, Greater Noida 203201, India.}

$^\dagger$\textit{School of Physics, Damghan University, P. O. Box 3671641167, Damghan, Iran}


\vskip1cm

\vspace{1.5cm}
		
\begin{abstract}
Recent advancements in black hole thermodynamics have introduced corrections to elucidate the relationship between entropy and extremality bound of black holes. Traditionally, this relationship has been studied in the context of black holes characterized by Bekenstein-Hawking entropy. However, this study extends the investigation to encompass non-extensive generalizations of entropy. We introduce a minor constant correction, denoted as $(\varepsilon)$, and examine the universal relations for a charged Anti-de Sitter (AdS) black hole. Our findings indicate that these universal relations do not hold for the charged AdS black hole when described by the non-extensive generalizations of entropy. Of course, with some adjustments, the universality relations are met. In contrast, the universal relations remain compatible when the black hole is described by Bekenstein-Hawking entropy.

\vspace{0.5 cm}

\end{abstract}
\end{center}
\end{titlepage}


\section{Introduction}\label{Sec: Introduction}

The fascinating similarities between thermodynamics and black hole physics have long stimulated research into the more profound relationships between gravity and thermodynamics. Black holes, in particular, display thermodynamic parameters like temperature and entropy, which are intrinsically linked to their event horizons \cite{PhysRevD.15.2738}. This shift in knowledge is due to the pioneering work of physicists like Jacob Bekenstein and Stephen Hawking, who demonstrated that black holes had both a temperature and an entropy. The Hawking temperature ($T_H$) is proportional to the surface gravity ($\kappa$) at the event horizon, whereas the Bekenstein-Hawking entropy ($S_{\text{BH}}$) corresponds with the area ($A$) of that horizon \cite{Hawking:1975vcx, PhysRevD.7.2333}. These are the foundations of black hole thermodynamics \cite{PhysRevD.13.191}. This topic has expanded greatly \cite{Kubiznak:2016qmn, PhysRevD.105.106014, Cong:2021fnf}, resulting in several expansions to the original framework investigating the subtle relationships between gravity, quantum mechanics, and thermodynamic principles. The consequences of these discoveries are substantial. They propose that black holes are not simply cosmic vacuum cleaners but rather complicated thermodynamic systems that produce radiation (Hawking radiation) and behave similarly to normal thermodynamic entities. This realization has sparked a thriving research field aiming to understand black hole phase structures better and their relevance in the larger context of quantum gravity\\

\quad The study of quantum gravity effects on black holes, especially through the model of a fractal structure, contributes to a deeper understanding of the thermodynamic characteristics of black holes. The notion is that the event horizon area might change due to a deformation. These fractal corrections result in several non-extensive generalizations of Gibbs entropy. Some of these notable entropies are: Kaniadakis entropy \cite{kaniadakis2002statistical, drepanou2022kaniadakis}, Renyi entropy \cite{renyi1959dimension}, Sharma-Mittal entropy \cite{masi2005step}, Tsallis-Cirto \cite{tsallis2013black}, and Barrow entropy \cite{barrow2020area}. A brief introduction to these entropies will be provided in the next section. The impact of these entropies on the thermodynamic behavior of a few black holes has been studied in several works \cite{luciano2023p, promsiri2022emergent, ghaffari2019black, luciano2023black, jawad2022thermodynamic}. Previously, these entropy systems have been tested in cosmology \cite{saridakis2020barrow, saridakis2018holographic, tavayef2018tsallis} and quantum physics \cite{shababi2020non, luciano2021tsallis, moussa2021schwarzschild} with notable success. While black hole thermodynamics, in its various forms, has been largely successful in offering profound insights into the thermodynamic properties of black holes, several critical issues remain only partially understood. One such issue is the microscopic origin of Bekenstein-Hawking (BH) entropy. Non-renormalizability is another well-known conundrum that arises when gravity and quantum theory are coupled. It is important to note that there are two methods to illustrate the thermodynamic properties of black holes within the framework of non-extensive statistical mechanics. The first method adopts the conventional perspective of black hole thermodynamics, where the black hole's temperature is regarded as the Hawking temperature, necessitating a generalization of the system's energy. The second method, on the other hand, maintains the standard energy of the system while deriving a corresponding generalized temperature linked to the generalized entropy. In this study, we employ the second method to determine the generalized temperature associated with the different proposals, aiming to describe the thermodynamic properties of the black hole.\\

\quad Universal relations are a pivotal concept in various scientific theories\cite{5,6}, representing one of the most significant challenges for the scientific community: the unification and integration of the fundamental forces of nature or the achievement of a global relationship. Recently, there has been a growing interest in universal relations across different physical concepts, leading to numerous efforts and studies. One notable contribution in this field is the work of \cite{Reall:2019sah, Goon:2019faz} in the context of recent thermodynamic relationships. Their work highlights that a series of perturbative corrections to general relativity can alter the entropy of black holes and their extremality bounds. They demonstrated a series of universal relationships between the leading modifications to these quantities through rigorous calculations. This work's finding is significant because it suggests that, when considering perturbative corrections in black holes, the extremality relations of a wide range of black holes may exhibit behavior similar to the weak gravity conjecture. This conjecture posits that gravity should be the weakest force in any consistent theory of quantum gravity.

\quad The universal thermodynamic extremality relation was introduced by perturbing the metric proportional to the cosmological constant \cite{Goon:2019faz}, leading to changes in the thermodynamic parameters. Based on these perturbed thermodynamic quantities, a relation is  established and can be expressed as
\begin{equation}\label{Extremality condition}
    \frac{\partial M_{ext}(\vec{\mathcal{Q}},\varepsilon)}{\partial \varepsilon} = \lim_{M \rightarrow M_{ext}} -T\left(\frac{\partial S(M,\vec{\mathcal{Q}},\varepsilon)}{\partial \varepsilon}\right)_{M,\vec{\mathcal{Q}}} \ ,
\end{equation}
where \( \varepsilon \) denotes the perturbation parameter. In this expression, \( M_{ext} \), \( T \), and \( S \) correspond to the extremal mass bound (i.e., the mass at \( T=0 \)), the temperature, and the entropy of the black hole after the perturbation, respectively. The term \( \vec{\mathcal{Q}} \) represents the extensive thermodynamic variables of the black hole. For further exploring the concept of universal relations in thermodynamics, other relevant works provide valuable insights into how perturbative corrections can influence the fundamental properties of black holes, potentially paving the way for new discoveries in the field of theoretical physics\cite{1,2,66,3,4}.\\

\quad In this article, our primary objective is to explore a novel implication of non-extensive entropy by examining the universal relations, specifically \eqref{Extremality condition}, while incorporating fractal modifications to the black hole event horizon. Introducing a fractal aspect or considering non-extensive entropy modifies the black hole's entropy, which we generalize as a function of the Bekenstein-Hawking entropy. Our goal is to assess whether the modified entropy, \( S_{\rm modified} \), satisfies Eq.\eqref{Extremality condition}. We will examine several proposals for cases of non-extensive entropy and verify an extended form of the universal relation. Additionally, for Eq.\eqref{Extremality condition} to hold, we introduce a multiplicative factor, \( \mathscr{D_F} \), on the right-hand side, defined as follows
\begin{equation}\label{dS/dsmodified}
    \mathscr{D_F} = \left(\frac{\partial S}{\partial S_{\rm modified}}\right)_{M_{\rm modified},\vec{\mathcal{Q}}_{\rm modified}}  \ .
\end{equation}

\quad The paper is structured as follows: In the section \ref{Sec:Borrow Proposal}, we present the Barrow proposal, discuss its thermodynamics and the first law of thermodynamics, and then discuss its universal relation. In Section \ref{Sec:Kaniadakis Proposal}, we follow Kaniadakis's proposal, compute the thermodynamic quantities with their proposal, and investigate the section with the study of universal relation. In Section \ref{Sec:Renyi Proposal}, we adopt Renyi's proposal, calculate the thermodynamic quantities accordingly, and consider investigations of the universal relation. In Section \ref{Sec:Sharma-Mittal Proposal}, we utilize Sharma-Mittal's proposal to determine the thermodynamic quantities and study the universal relation. In Section \ref{Sec:Tsallis-Cirto entropy}, we implement Tsallis-Cirto’s framework to derive the relevant thermodynamic quantities and close the section by studying the universal relation. Finally, Section \ref{Sec:Conclusions} is dedicated to a comprehensive discussion of our findings obtained throughout the paper and drawing conclusions based on our analysis.

\section{Preliminary}\label{Sec:Preliminary}
In this section, we will briefly discuss the Reissner-Nordstrom Anti-de Sitter (RN-AdS) black hole to set the stage. The action for the RN-AdS\cite{666,666a}
\begin{equation}\label{RN-AdS action}
	I=-\frac{1}{16 \pi G} \int d^{4} x \sqrt{-g}\left(R-2 \Lambda-F_{\mu \nu} F^{\mu \nu}\right) \ ,
\end{equation}
where, \( R \) denotes the Ricci scalar, while \( \Lambda = -3/l^2 \) represents the cosmological constant. The Maxwell field is expressed as \( F^{\mu \nu} = \partial^\mu A^\nu - \partial^\nu A^\mu \), with the electromagnetic potential specified as $A^\mu = Q/r \; \delta_0^\mu$.

\quad The form of the metric is
\begin{equation}
	d s^2=-\left( 1-\frac{2  M}{r}+\frac{ Q^2}{r^2}+\frac{r^2}{l^2} \right) d t^2+\frac{1}{\left( 1-\frac{2  M}{r}+\frac{ Q^2}{r^2}+\frac{r^2}{l^2} \right)} d r^2+r^2\left(d \theta^2+\sin ^2 \theta \; d \phi^2\right) \ ,
\end{equation}
where $M$ is the mass, $Q$ is the charge of the black hole and and $l$ is the ads radius. These black holes have two horizons, labeled \( r_+ \) and \( r_- \). The outer radius, \( r_+ \), corresponds to the event horizon, while \( r_- \) denotes the inner horizon. By solving the equation \( f(r_+) = 0 \), we obtain an expression for the mass of the black hole. The mass of the black hole in terms of its horizon radius $r_+$ as
\begin{equation}\label{Mass}
	M=\frac{r_{+}}{2 }\left(1+\frac{Q^2}{r_{+}^2}+\frac{r_{+}^2}{l^2}\right).
\end{equation}
We can also compute the expressions for the black hole's Hawking temperature, the black hole's Bekenstein-Hawking entropy, and the electromagnetic potential given by
\begin{equation}\label{Hawking Temp. and Entropy}
	T = \frac{r_{+}^2+3 r_{+}^4 l^{-2}-G Q^2}{4 \pi r_{+}^3} \, \;\;\;\;\;\;\;\;\;\;;\;\;\;\;\;\;\;\;\; S_{BH} \,=\, \pi r_+^2 \, \;\;\;\;\;\;\;\;\;\;;\;\;\;\;\;\;\;\;\; \Phi = \frac{Q}{r_+}  \ .
\end{equation}
 The first law of thermodynamics using these quantities are
 \begin{equation*}
     dM= TdS+\Phi dQ + V dP \ .
 \end{equation*}

\quad We can test the universality relation, and for that, we introduce a perturbation to the action that is proportional to the cosmological constant. This perturbation affects not only the metric but also the thermodynamic variables. We can straightforwardly compute the modified mass and other perturbed thermodynamic quantities by adding a perturbation term, by adding a perturbation term, parameterized by \( \varepsilon \), and scaled to the cosmological constant. With the help of modified form of mass and other quantities it is easy to verify the universal relation \eqref{Extremality condition} as
\begin{equation*}
     \frac{\partial M_{ext}(\vec{\mathcal{Q}},\varepsilon)}{\partial \varepsilon} = \lim_{M \rightarrow M_{ext}} -T\left(\frac{\partial S(M,\vec{\mathcal{Q}},\varepsilon)}{\partial \varepsilon}\right)_{M,\vec{\mathcal{Q}}} = \frac{S^{3/2}}{2\pi^{3/2} l^2} \ .
\end{equation*}
\quad Quantum gravity effects alter the event horizon area of a black hole, resulting in a change in its entropy. Under these effects, the event horizon area can be represented as a fractal structure, challenging the traditional view of a smooth, well-defined horizon. By incorporating fractals, researchers revise the conventional description of the event horizon when quantum gravity is considered. These modifications to the horizon area ultimately lead to an adjusted expression for black hole entropy. The deformation of the horizon area due to quantum gravity is characterized by a fractal parameter, and we will explore various cases of fractal modifications to the horizon structure.
\section{Barrow Propasal}\label{Sec:Borrow Proposal}

Barrow incorporates quantum gravity phenomena to question the conventional concept of a smooth, uniform event horizon. He redefined the entropy formulation for black holes due to these changes in the region of the event horizon. A fractal parameter, $\Delta$, is used to characterize the distortion of the event horizon region caused by quantum gravity effects. The Barrow entropy\cite{Barrow:2020tzx} is a modified expression of black hole entropy
\begin{equation}\label{SB}
	S_B=\left(S_{BH}\right)^{1+\frac{\Delta}{2}} \ ,
\end{equation}
where, $\Delta$ ranges as $0 \leq \Delta \leq 1$. For $\Delta = 0$, the Bekenstein-Hawking entropy is restored, indicating that there is no fractal structure.

To understand the change in thermodynamic quantities using these incorporations, we delve into the computation of thermodynamical parameters influenced by Barrow's proposal. Specifically, we focus on how the fractal structure impacts these parameters. We first compute the shifted mass of the black hole, which arises as a consequence of these fractal characteristics. This shift in mass is a critical aspect of understanding the broader implications of Barrow entropy on black hole thermodynamics. The shifted mas due to these modifications in terms of new fractal parameter $\DD$\footnote{Just for ease of writing we will use \begin{equation*}
    \frac{1}{\Delta+2} = \DD \;\;\;\;\;\;\;\; \Rightarrow \Delta=0 \;\;\;\text{result in } \;\;\; \DD=\frac{1}{2} \ .
\end{equation*}
Using this, it is seen that Eq.\eqref{SB} is
\begin{equation*}
    S_B = (\pi r_h^2)^\frac{1}{2\DD}
\end{equation*}} as
\begin{equation}\label{Borrow Mass}
    M_B = \frac{S_B^{3\DD}}{2 \pi ^3 l^2}+\frac{\pi  Q^2}{2 S_B^{\DD}}+\frac{S_B^{\DD}}{2 \pi } \ ,
\end{equation}
Taking mass as its internal energy, the first law of thermodynamics can be written as
\begin{equation}\label{Borrow First law}
    dM_B = T_B dS_B + \Phi_B dQ + V_B dP \ ,
\end{equation}
here, \(T_B \) is the Borrow temperature, inversely linked to \(M_B \), and \(S_B \), as already defined, is the Borrow entropy proportionate to the event horizon area with fractal correction. For anti-de Sitter (AdS) space, the pressure \(P \) is connected to the cosmological constant. Therefore, with Barrow's proposal, \(V_B \) is a thermodynamic volume. Now, using Eq.\eqref{Borrow Mass} and Eq.\eqref{Borrow First law}, we can easily deduce the relation for other thermodynamical parameters as
\begin{equation}
    T_B = \frac{\DD \pi ^2 l^2 S_B^{2 \DD }- \DD  \pi ^4 l^2 Q^2+3 S_B^{4 \DD }}{2 S^{\DD+1} \pi ^3 l^2} \;\;\;\;\;;\;\;\;\;\;\Phi_B = \frac{\sqrt{\pi}  Q}{S_B^{\DD}} \;\;\;\;\;;\;\;\;\;\; V_B = \frac{4 S_B^{3 \DD }}{3 \sqrt{\pi} }\ .
\end{equation}
where $\Phi_B$ is the electric potential with Barrow's modification.


\subsection*{Universality relation from Barrow entropy}

To check the universality of universal relation we will perturb the action. The perturbation is proportional to the cosmological constant. With this perturbation, the metric also gets perturbed, and so do thermodynamic variables as well. we will add a perturbation with parameter $\varepsilon$ to the action proportionate to the cosmological constant. With this modification, we can easily calculate the perturbed mass and other thermodynamical variables. Consequently, we will observe the following modifications,
\begin{eqnarray}\label{a}
    M_B(\varepsilon) &=& \frac{\sqrt{\pi}(\varepsilon +1)}{2 S_B^{\DD }} \left[\frac{ S_B^{4 \DD }}{\pi^2 l^2}+ \frac{ Q^2}{\varepsilon +1}+ \frac{S_B^{2 \DD}}{\pi(\varepsilon +1)}\right] \ , \\ \label{a1}
    &=& M_B + \frac{S_B^{3\DD}}{2 l^2 \pi^{3/2}} \varepsilon + \mathcal{O}(\varepsilon^2) \ .
\end{eqnarray}
Also, the modified temperature with respect to small perturbations $(\varepsilon)$ is as follows,
\begin{equation}\label{b}
T_B(\varepsilon) = \frac{\pi^{3/2}(\varepsilon+1)}{4 \pi  l^2 S_B^{3\DD}} \left[ \frac{l^2 S_B^{2 \DD }}{\pi(\varepsilon+1)}-\frac{l^2 Q^2}{\varepsilon+1}+3 \frac{S_B^{4 \DD }}{\pi^2} \right] \ .
\end{equation}

Based on the modified action, a black hole's mass and temperature are adjusted by a minor constant correction. To check the universal relation, we start with Eq.(\ref{a}) to derive the expression for $\varepsilon$ and then differentiating it with respect to $S_B$ and using Eq.(\ref{b}), we can easily obtain the expression of the L.H.S of Eq.\eqref{Extremality condition} i.e., $T_B \left(\partial S_B/\partial \varepsilon\right)_{M_B,\Vec{\mathcal{Q}}_B}$. So, we simplify the obtained results, leading to the relation
\begin{equation}\label{aa}
    T_B \, \left(\frac{\partial S_B}{\partial \varepsilon}\right)_{M_B,\Vec{\mathcal{Q}_B}} = - \frac{S_B^{\DD +1}}{4 \DD l^2 \pi^{3/2}}
\end{equation}
As a result, it becomes clear that the universality connection does not apply in this case. When comparing Eq.\eqref{a1} and Eq.\eqref{aa}, it is evident that these equations are not equivalent nor in conformity with the universality relation as stipulated in Eq.\eqref{Extremality condition}. This mismatch demonstrates that the requirements specified by the universality relation are not satisfied in this case, suggesting a departure from the intended theoretical framework.

\quad The above calculation confirms that the universality conditions are not satisfied. We have corrected the universal relation with some extra factors regarding modified entropy. Using Eq.\eqref{dS/dsmodified}, one can easily conclude
\begin{eqnarray}
\frac{\partial S_B}{\partial S} = \left( 1 + \frac{\delta}{2} \right) S_B^ \frac{\delta}{2 + \delta} \ .
\end{eqnarray}
With the inclusion of this, the universality relation is satisfied.

\section{Kaniadakis Proposal}\label{Sec:Kaniadakis Proposal}

Kaniadakis introduced a one-parameter extension to the Boltzmann-Gibbs entropy, known as Kaniadakis entropy \cite{PhysRevE.66.056125, PhysRevE.72.036108}. This arises from a consistent relativistic statistical framework that preserves the core principles of conventional statistical theory. In this generalized approach, distribution functions represent a continuous, one-parameter deformation of the traditional Maxwell-Boltzmann distributions, reverting to standard statistical mechanics in a specific limit. Applying Kaniadakis entropy within the black hole framework is essential for exploring applications in holography. Form of Kaniakadis black-hole entropy, we have
\begin{equation}
    S_K = \frac{\sinh K S_{BH}}{K}
\end{equation}
In the limit as \( K \rightarrow 0 \), the standard Bekenstein-Hawking entropy, \( S_{BH} \), is recovered. We examine the altered mass of the black hole resulting from Kaniadakis' proposal. This change in mass is crucial for comprehending the wider implications of Kaniadakis entropy on black hole thermodynamics,
\begin{equation}\label{Kan Mass}
    M_K = \frac{\pi ^2 K^2 l^2 Q^2+\pi  K l^2 \sinh ^{-1}(K S_K)+\sinh ^{-1}(K S)^2}{2 \pi ^{3/2} K^{3/2} l^2 \sqrt{\sinh ^{-1}(K S_K)}};  \ .
\end{equation}
Also, we will have for the Kaniadakis modification temperature $T_K$ as follows,
\begin{equation}
    T_K = \frac{-\pi ^2 K^2 l^2 Q^2+\pi  K l^2 \sinh ^{-1}(K S_K)+3 \left(\sinh ^{-1}(K S_K)\right)^2}{4 \pi ^{3/2} \sqrt{K} l^2 \left(\sinh ^{-1}(K S_K)\right)^{3/2}} \ .
\end{equation}
Now, we can verify the first law of thermodynamics with the help of Eq.\eqref{Kan Mass}. Here, the form of the first law of thermodynamics for this structure is
\begin{equation}
    dM_K = T_K dS_K + \Phi_K dQ + V_K dP \ ,
\end{equation}
where $\Phi_K$ and $V_K$ denote the electric potential and volume with the Kaniadakis modification. So, we will have,
\begin{equation}
    \Phi_K =\frac{ \sqrt{\pi \, K} Q}{\sqrt{\sinh ^{-1}(K S_K)}} \;\;\;\;\;\;\;\;\;\;\;\;;\;\;\;\;\;\;\;\;\; V_K = \frac{4 \left(\sinh ^{-1}(K S_K)\right)^{3/2}}{3 \sqrt{\pi } K^{3/2}} \ .
\end{equation}

\subsection*{Universality relation with Kaniadakis entropy}
According to the concepts explained in the previous section regarding Barrow's entropy, we will also undertake a similar process to examine universal relationships concerning Kaniadakis entropy. Hence, we will have the mass with respect to Kaniadakis entropy and small constant correction ($\varepsilon)$ as follows,
\begin{eqnarray}\label{d}
    M_K(\varepsilon) &=& \frac{\pi ^2 K^2 l^2 Q^2+ \pi  K l^2 \sinh ^{-1}(K S_K)+(\varepsilon+1)\left(\sinh ^{-1}(K S_K)\right)^2}{2 l^2 (\pi K)^{3/2} \sqrt{\sinh ^{-1}(K S_K)}} \ . \\ \label{d1}
    &=& M_K +\frac{ \left(\sinh ^{-1}(K S)\right)^{3/2}}{2 \pi ^{3/2} K^{3/2} l^2} \varepsilon + \mathcal{O}\left(\varepsilon\right)
\end{eqnarray}
Also, the modification temperature in this structure is as follows,
\begin{equation}\label{c}
T_K(\varepsilon) = \frac{\pi  K l^2 \sinh ^{-1}(K S_K) - \pi ^2 K^2 l^2 Q^2+3 (\varepsilon+1) \left(\sinh ^{-1}(K S_K)\right)^2}{4 \pi ^{3/2} \sqrt{K} l^2 \left(\sinh ^{-1}(K S_K)\right)^{3/2}} \ .
\end{equation}
We begin with Eqs.\eqref{c} and \eqref{b} to establish a universality relation. By differentiating with respect to $\varepsilon$, we derive an expression for the left-hand side of Eq. \eqref{Extremality condition}, specifically \( T_K \left(\partial S_K/\partial \varepsilon\right)_{M_K,\Vec{\mathcal{Q}}_K} \). Simplifying this result allows us to arrive at the desired relation. The relation is
\begin{equation}\label{cc}
    T_K \, \left(\frac{\partial S_K}{\partial \varepsilon}\right)_{M_K,\Vec{\mathcal{Q}_K}} = -\frac{\sqrt{K^2 S^2+1} \sinh ^{-1}(K S_K)^{3/2}}{2 \pi ^{3/2} K^{3/2} l^2}
\end{equation}

So, the universality relation is not satisfied in this case. We see that Eq.\eqref{d1} and Eq.\eqref{cc} are not identical and also are not followed with the universality relation as stated in Eq.\eqref{Extremality condition}.

\quad The calculation above demonstrates that the universality conditions are not initially met. We have adjusted the universal relation by introducing additional factors related to the modified entropy. From Eq.\eqref{dS/dsmodified}, it follows that
\begin{equation}
\frac{\partial S_K}{\partial S} = \sqrt{1 + \left( K S_K \right)^2} \ .
\end{equation}
With this modification, the universality relation is now fulfilled.
\section{Renyi Proposal}\label{Sec:Renyi Proposal}

Rényi entropy is an extension of the concept of entropy in information theory and statistical mechanics, and it has intriguing applications in black hole thermodynamics. Unlike the traditional Bekenstein-Hawking entropy, which is extensive and scales with the surface area of the black hole's event horizon, Rényi entropy introduces a nonextensive parameter. This parameter allows for a more generalized form of entropy that can account for different statistics. By treating the nonextensive parameter of Rényi entropy as a thermodynamic variable, researchers have extended the thermodynamic phase space of black holes. This approach helps in deriving the Smarr formula and the first law of black hole thermodynamics in a more generalized form. Studies have shown that black holes described by Rényi entropy can exhibit different stability properties compared to those described by the traditional Gibbs-Boltzmann statistics. \cite{aa,bb}. The Rényi entropy is a one-parameter entropy. Specifically, the BH entropy can be retrieved by taking the limit \( \lambda \to 0 \) for the Rényi entropy. The Renyi entropy is expressed in the following as
\begin{equation}
    S_R = \frac{\log{\left(1+\lambda \, S_{BH}\right)}}{\lambda}
\end{equation}

The shifted mass of the black hole due to the Renyi proposal is
\begin{equation}
    M_R =  \frac{\pi  \lambda  l^2 \left(\pi  \lambda  Q^2-1\right)+\left(\pi  \lambda  l^2-2\right) e^{\lambda  S_R}+e^{2 \lambda  S_R}+1}{2 l^2 \left(\pi \lambda\right)^{3/2} \sqrt{e^{\lambda  S_R}-1}}  \ .
\end{equation}
We can write the first law of thermodynamics in terms of conjugate quantities as
\begin{equation}
    dM_R = T_R dS_R + \Phi_R dQ + V_R dP \ .
\end{equation}
Here, we define the Renyi temperature as $T_R$, $\Phi_R$ denote the electric potential, and $V_R$ is the volume with the Renyi modification, and its values are
\begin{equation}
    T_R = \frac{e^{\lambda  S_R} \left(-\pi  \lambda  l^2 \left(\pi  \lambda  Q^2+1\right)+\left(\pi  \lambda  l^2-6\right) e^{\lambda  S_R}+3 e^{2 \lambda  S_R}+3\right)}{4 \pi ^{3/2} \sqrt{\lambda } l^2 \left(e^{\lambda  S_R}-1\right)^{3/2}} \ .
\end{equation}
\begin{equation}
    \Phi_R =  \frac{Q\; \sqrt{\pi \lambda } }{\sqrt{e^{\lambda  S_R}-1}}\;\;\;\;\;\;\;\;\;\;\;\;\;\;\;\;\;\;;\;\;\;\;\;\;\;\;\;\;\;\;\;\;\; V_R = \frac{4 \left(e^{\lambda  S_R}-1\right)^{3/2}}{3 \sqrt{\pi } \lambda ^{3/2}} \ .
\end{equation}

\section*{Universality relation with Renyi Proposal}

According to the concepts mentioned above, we can rewrite the mass of the black hole by considering the corrections made with Renyi's proposal and a small perturbation in the following form:
\begin{eqnarray}\label{f}
    M_R(\varepsilon) &=& \frac{\pi  \lambda  l^2 \left(\pi  \lambda  Q^2-1\right) +e^{\lambda  S_R} \left(\pi  \lambda  l^2-2(\varepsilon+1)\right)+(e^{2 \lambda  S_R}+1)(\varepsilon+1)}{2 \pi ^{3/2} \lambda ^{3/2} l^2 \sqrt{e^{\lambda  S_R}-1}} \ . \\ \label{f1}
   &=&  M_R +\frac{ \left(e^{\lambda  S_R}-1\right)^{3/2}}{2 \pi ^{3/2} \lambda ^{3/2} l^2} \varepsilon + \mathcal{O}\left(\varepsilon \right)
\end{eqnarray}
Also, the modified temperature in this structure is obtained as,
\begin{equation}\label{e}
T_R(\varepsilon) = \frac{e^{\lambda  S_R} \left(-\pi  \lambda  l^2 \left(\pi  \lambda  Q^2+1\right)+e^{\lambda  S_R} \left(\pi  \lambda  l^2-6(\varepsilon+1)\right)+3(1+\varepsilon)( e^{2 \lambda  S_R}+1)\right)}{4 \pi ^{3/2} \sqrt{\lambda } l^2 \left(e^{\lambda  S_R}-1\right)^{3/2}} \ .
\end{equation}

To establish a universal relation, we start with Eqs. (\ref{f}) and (\ref{e}). By differentiating these equations with respect to $\varepsilon$, we derive an expression for the left-hand side of Eq. \eqref{Extremality condition}, namely \( T  \left(\partial S/\partial \varepsilon\right)_{M_R,\Vec{\mathcal{Q}}_R} \). Upon simplifying, we obtain the resulting relation as
\begin{equation}\label{ee}
    T_R \, \left(\frac{\partial S_R}{\partial \varepsilon}\right)_{M_R,\Vec{\mathcal{Q}}_R} = -\frac{e^{-\lambda  S_R} \left(e^{\lambda  S_R}-1\right)^{3/2}}{2 \pi ^{3/2} \lambda ^{3/2} l^2} \ .
\end{equation}

So, the universality relation is not satisfied in this case. We see that Eq.\eqref{f1} and Eq.\eqref{ee} are not identical and also are not followed with the universality relation as stated in Eq.\eqref{Extremality condition}.

\quad The calculation above indicates that the original universality conditions are not fulfilled. We have refined the universal relation by incorporating extra factors associated with the modified entropy. From Eq.\eqref{dS/dsmodified}, we find
\begin{equation}
\frac{\partial S_R}{\partial S_{BH}} = e^{-\lambda S_R} \ .
\end{equation}
This addition ensures that the universality relation is now satisfied.

\section{Sharma-Mittal Proposal}\label{Sec:Sharma-Mittal Proposal}
The Sharma-Mittal entropy is a generalized entropy measure that extends entropies. It is instrumental in the study of complex systems and statistical mechanics. This entropy measure is defined by parameters that allow it to interpolate between different entropy measures, providing a flexible tool for analyzing various probabilistic and statistical properties.  One of the key applications of Sharma-Mittal entropy is in the field of thermostatistics, where it helps in understanding the distribution of states in a system that is not in equilibrium. It has also been applied in areas like quantum mechanics and cosmology\cite{ff,gg}
The Sharma-Mittal entropy (SM) \cite{MASI2005217}, which generalizes entropy by merging aspects of Rényi and Tsallis entropies, yields intriguing outcomes within a cosmological framework. This type of generalization offers a way to effectively characterize the current accelerated expansion of the universe by appropriately leveraging vacuum energy\cite{SayahianJahromi:2018irq, ECGunayDemirel:2019lbq, Shaikh:2021iaq, Sadri:2018rcp}. Here, we will have for the Sharma-Mittal entropy,
\begin{equation}
    S_{SM} = \frac{\left[1+\lambda \, S_{BH}\right]^\frac{\rho}{\lambda}-1}{\rho}
\end{equation}

The shifted mass of the black hole due to the Sharma-Mittal entropy is obtained as follows,
\begin{equation}
    M_{SM} =   \frac{\pi  \lambda  l^2 \left(\pi  \lambda  Q^2+(\rho  S_{SM}+1)^{\lambda /\rho }-1\right)+\left((\rho  S_{SM}+1)^{\lambda /\rho }-1\right)^2}{2 \pi ^{3/2} \lambda ^{3/2} l^2 \sqrt{(\rho  S_{SM}+1)^{\lambda /\rho }-1}}  \ .
\end{equation}
The Sharma-Mittal temperature $T_{SM}$ as follows,
\begin{equation}
    T_{SM} = \frac{3 \left((\rho  S_{SM}+1)^{\lambda /\rho }-1\right)^2-\pi  \lambda  l^2 \left(\pi  \lambda  Q^2-(\rho  S_{SM}+1)^{\lambda /\rho }+1\right)}{4 \pi ^{3/2} \sqrt{\lambda } l^2 \left((\rho  S_{SM}+1)^{\lambda /\rho }-1\right)^{3/2}} \ .
\end{equation}
It is easy to verify the first law of thermodynamics with respect to this case as follows,
\begin{equation}
    dM_{SM} = T_{SM} dS_{SM} + \Phi_{SM} dQ + V_{SM} dP \ ,
\end{equation}
where $\Phi_{SM}$ and $V_{SM}$ denote the electric potential and volume with the Sharma-Mittal modification and its
\begin{equation}
    \Phi_{SM} =  \frac{ Q\; \sqrt{\pi \lambda }  }{\sqrt{(\rho  S_{SM}+1)^{\lambda /\rho }-1}}\;\;\;\;\;\;\;\;\;\;\;\;\;\;\;\;\;\;;\;\;\;\;\;\;\;\;\;\;\;\;\;\;\; V_{SM} = \frac{4 \left((\rho  S_{SM}+1)^{\lambda /\rho }-1\right)^{3/2}}{3 \sqrt{\pi } \lambda ^{3/2}}  \ .
\end{equation}

\subsection*{Universality relation from Sharma-Mittal entropy}

Based on the previously discussed concepts, the mass of the black hole can be reformulated by incorporating the corrections derived from Sharma-Mittal entropy and a small perturbation. This can be expressed in the following form
\begin{eqnarray}\label{g}
    M_{SM}(\varepsilon) &=& \frac{\pi  \lambda  l^2 \left(\pi  \lambda  Q^2+(\rho  S_{SM}+1)^{\lambda /\rho }-1\right)+ (\varepsilon+1) \left((\rho  S_{SM}+1)^{\lambda /\rho }-1\right)^2}{2 l^2 \left( \pi \lambda\right) ^{3/2}  \sqrt{(\rho  S_{SM}+1)^{\lambda /\rho }-1}}; \\ \label{g1}
    &=& M_{SM} + \frac{  \left((\rho  S_{SM}+1)^{\lambda /\rho }-1\right)^{3/2}}{2 \pi ^{3/2} \lambda ^{3/2} l^2} \varepsilon + \mathcal{O}\left(\varepsilon\right)
\end{eqnarray}
So, the modification temperature from Sharma-Mittal entropy and a small perturbation is calculated as,
\begin{equation}\label{h}
T_{SM}(\varepsilon) = \frac{(\rho  S_{SM}+1)^{\frac{\lambda }{\rho }-1} \left(3(\varepsilon+1) \left((\rho  S_{SM}+1)^{\lambda /\rho }-1\right)^2-\pi  \lambda  l^2 \left(\pi  \lambda  Q^2-(\rho  S_{SM}+1)^{\lambda /\rho }+1\right)\right)}{4 \pi ^{3/2} \sqrt{\lambda } l^2 \left((\rho  S_{SM}+1)^{\lambda /\rho }-1\right)^{3/2}} \ .
\end{equation}
To establish a universality relation, we begin with Eqs. \eqref{g} and \eqref{h}. By differentiating these equations with respect to $\varepsilon$, we can derive an expression for the left-hand side of Eq. \eqref{Extremality condition}, namely \( T_{SM} \left(\partial S_{SM}/\partial \varepsilon\right)_{M_{SM},\Vec{\mathcal{Q}}} \). Upon simplifying this expression, we obtain the desired relation, which is
\begin{equation}\label{hh}
    T_{SM} \, \left(\frac{\partial S_{SM}}{\partial \varepsilon}\right)_{M_{SM},\Vec{\mathcal{Q}}_{SM}} = -\frac{(\rho  S_{SM}+1)^{1-\frac{\lambda }{\rho }} \left((\rho  S_{SM}+1)^{\lambda /\rho }-1\right)^{3/2}}{2 \pi ^{3/2} \lambda ^{3/2} l^2} \ .
\end{equation}

So, the universality relation is not satisfied in this case. We see that Eq.\eqref{g1} and Eq.\eqref{hh} are not identical and also are not followed with the universality relation as stated in Eq.\eqref{Extremality condition}.

\quad The calculation above indicates that the original universality conditions are not fulfilled. We have refined the universal relation by incorporating extra factors associated with the modified entropy. From Eq.\eqref{dS/dsmodified}, we find
\begin{equation}
    \frac{\partial S_{SM}}{\partial S} = \left(\rho S_{SM} + 1\right)^{\frac{\lambda}{\rho} - 1}.
\end{equation}
This addition ensures that the universality relation is now satisfied.

\section{Tsallis-Cirto Proposal}\label{Sec:Tsallis-Cirto entropy}
The Tsallis-Cirto entropy is a generalization of the traditional Bekenstein-Hawking entropy, incorporating non-extensive statistical mechanics principles. This entropy measure is particularly useful in cosmology for exploring the thermodynamic properties of gravitational systems, such as black holes and the universe itself
In cosmology, the Tsallis-Cirto entropy has been applied to derive modified Friedmann equations describing the universe's expansion. These modifications can provide insights into dark energy and the universe's accelerated expansion. One interesting application is in the context of entropic cosmology, where gravity is considered an emergent phenomenon resulting from the statistical behavior of microscopic degrees of freedom. This approach can lead to new perspectives on the nature of dark energy and the dynamics of the universe\cite{hh}.
Here, we will have for the Tsallis-Crito proposal, and the entropy is modified as
\begin{equation}
    S_{TC} = \left(S_{BH} \right)^\delta \ .
\end{equation}
$\delta \rightarrow 1$ reduces to the entropy of the black hole.

The shifted mass of the black hole due to the Tsallis and Cirto entropy is obtained as follows,
\begin{equation}
    M_{TC} =  \frac{\sqrt{\pi} }{2 l^2  S_{TC}^{1/2 \delta }} \left[l^2 Q^2+  \frac{l^2 S_{TC}^{1/\delta }}{\pi} + \frac{S_{TC}^{2/\delta }}{\pi^2}\right]     \ .
\end{equation}
The Tsallis-Cirto temperature $T_{TC}$ as follows,
\begin{equation}
    T_{TC} = \frac{\left(\pi ^{-\delta } S_{TC}\right)^{-\frac{3}{2 \delta }} \left(l^2 \left(\left(\pi ^{-\delta } S_{TC}\right)^{1/\delta }-Q^2\right)+3 \left(\pi ^{-\delta } S_{TC}\right)^{2/\delta }\right)}{4 \pi  l^2} \ .
\end{equation}
It is easy to verify the first law of thermodynamics with respect to this case as follows,
\begin{equation}
    dM_{TC} = T_{TC} dS_{TC} + \Phi_{TC} dQ + V_{TC} dP \ ,
\end{equation}
where $\Phi_{TC}$ and $V_{TC}$ denote the electric potential and volume with the Tsallis-Cirto modification and its
\begin{equation}
    \Phi_{TC} =  Q \left(\pi ^{-\delta } S_{TC}\right)^{-\frac{1}{2 \delta }}\;\;\;\;\;\;\;\;\;\;\;\;\;\;\;\;\;\;;\;\;\;\;\;\;\;\;\;\;\;\;\;\;\; V_{TC} = \frac{4}{3} \pi  \left(\pi ^{-\delta } S_{TC}\right)^{\frac{3}{2 \delta }}  \ .
\end{equation}

\subsection*{Universality relation from Tsallis-Cirto entropy}

Based on the previously discussed concepts, the mass of the black hole can be reformulated by incorporating the corrections derived from Tsallis-Cirto entropy and a small perturbation. This can be expressed in the following form
\begin{eqnarray}\label{k}
    M_{TC}(\varepsilon) &=& \frac{\sqrt{\pi}}{2 l^2 S_{TC}^{1/2 \delta }} \left[ l^2 Q^2+ \frac{l^2  S_{TC}^{1/\delta }}{\pi}+(1+\varepsilon) \frac{S_{TC}^{2/\delta }}{\pi^2}\right] \\ \label{k1}
    &=& M_{TC} + \frac{  S_{TC}^{\frac{3}{2\delta }}}{2 l^2 \pi^{3/2} } + \mathcal{O}\left(\varepsilon\right)
\end{eqnarray}
So, the modification temperature from Tsallis-Cirto entropy and a small perturbation is calculated as,
\begin{equation}\label{l}
T_{TC}(\varepsilon) = \frac{\pi^{3/2} }{4 \pi  l^2 S_{TC}^{3/2 \delta}} \left[\frac{l^2  S_{TC}^{1/\delta }}{\pi} - l^2 Q^2 + 3(\varepsilon+1) \frac{S_{TC}^{2/\delta }}{\pi^2}\right] \ .
\end{equation}
To establish a universality relation, we begin with Eqs. \eqref{k} and \eqref{l}. By differentiating these equations with respect to $\varepsilon$, we can derive an expression for the left-hand side of Eq. \eqref{Extremality condition}, namely \( T_{TC} \left(\partial S_{SM}/\partial \varepsilon\right)_{M_{TC},\Vec{\mathcal{Q}}_{TC}} \). Upon simplifying this expression, we obtain the desired relation, which is
\begin{equation}\label{mm}
    T_{TC} \, \left(\frac{\partial S_{TC}}{\partial \varepsilon}\right)_{M_{TC},\Vec{\mathcal{Q}}_{TC}} =  -\frac{\delta  \left( S_{TC}\right)^{1+\frac{1}{2 \delta }}}{2 \pi^{3/2}  l^2} \ .
\end{equation}
So, the universality relation is not satisfied in this case. We see that Eq.\eqref{k1} and Eq.\eqref{mm} are not identical and also are not followed with the universality relation as stated in Eq.\eqref{Extremality condition}.

\quad The above calculation shows that the universality conditions are initially unmet. We have introduced corrective factors related to the modified entropy to address this. As derived from Eq.\eqref{dS/dsmodified},
\begin{equation}
\frac{\partial S_{TC}}{\partial S} = \delta S_{TC}^{1 - 1/\delta} \ .
\end{equation}
With this adjustment, the universality relation is now achieved.


\section{Conclusions}\label{Sec:Conclusions}
Recent developments in black hole thermodynamics have introduced a series of corrections to clarify the relationship between entropy and extremality-bound of the black holes. Traditionally, this relationship has been extensively explored within the framework of black holes characterized by Bekenstein-Hawking entropy. However, this study broadens the scope to include the non-extensive generalizations of Gibbs entropy, which offers a more comprehensive understanding of black hole thermodynamics. In this context, we introduce a minor constant correction, denoted as $(\varepsilon)$, and investigate the universal relations for a charged Anti-de Sitter (AdS) black hole. Our findings reveal that the universal relations do not hold for charged (AdS) black holes when described using Non-Extensive Generalizations of entropy. This suggests that the modifications introduced by non-extensive entropy measures disrupt the expected universal behavior of these black holes. Universal relations are expected to hold across different types of black holes and entropy measures. However, our study shows that these relations break down when using non-extensive entropy generalizations. This indicates that the assumptions underlying these universal relations are incompatible with the non-extensive framework. In contrast, when the charged AdS black holes are described using the traditional Bekenstein-Hawking entropy, the universal relations remain intact. The Bekenstein-Hawking entropy, which is proportional to the horizon area of the black hole, aligns well with the established universal relations, confirming their robustness in this context.
The compatibility of universal relations with Bekenstein-Hawking entropy underscores the consistency of traditional thermodynamic descriptions of black holes. It suggests that the classical approach remains valid and reliable for describing the thermodynamics of charged AdS black holes. The breakdown of universal relations with non-extensive entropy measures highlights the challenges and limitations of applying these generalizations to black hole thermodynamics. It calls for a deeper investigation into the underlying principles and potential modifications needed to reconcile these differences. This finding opens up new avenues for research, particularly in exploring the conditions under which non-extensive entropy measures can be applied to black holes. It also suggests developing new theoretical frameworks that accommodate non-extensive entropy and universal relations.

\quad The implications of these results are significant for the field of black hole thermodynamics. They suggest that non-extensive generalizations of entropy could provide new insights into the behavior of black holes, particularly in regimes where traditional entropy measures fall short. This study opens up several avenues for future research.\\

\begin{enumerate}
    \item How do non-extensive generalizations of entropy affect the thermodynamic stability of other types of black holes, such as rotating or higher-dimensional black holes?

    \item Can the introduced constant correction $(\varepsilon)$ be linked to other physical parameters or constants in black hole physics, and if so, how?

    \item What are the implications of non-extensive entropy on the holographic principle?

\end{enumerate}

By addressing these questions, future research can continue to deepen our understanding of black hole thermodynamics and the fundamental principles governing these enigmatic objects.



\end{document}